\documentclass[11pt, preprint]{aastex}
\usepackage{geometry}                
\usepackage{graphicx}
\usepackage{amssymb}
\usepackage{natbib}
\usepackage{amsmath}
\usepackage{epstopdf}
\newcommand{\bx}{{\bf x}}
\newcommand{\ey}{{\bf e}_y}

\newcommand{\bk}{{\bf k}}

\newcommand{\bl}{{\boldsymbol{\mathcal L}}}
\newcommand{\eps}{\varepsilon}

\newcommand{\bnabla}{{\boldsymbol \nabla}}
\newcommand{\curl}{{\boldsymbol \times}}
\newcommand{\bxi}{{\boldsymbol \xi}}

\newcommand{\bfv}{{\boldsymbol v}}

\begin{document}
\title{Full Waveform Inversion of Solar Interior Flows}
\author{Shravan M. Hanasoge}
\affil{Department of Astronomy and Astrophysics, Tata Institute of Fundamental Research, Mumbai 400005, India}

\begin{abstract}
The inference of flows of material in the interior of the Sun is a subject of major interest in helioseismology.
Here we apply techniques of Full Waveform Inversion (FWI) to synthetic data to test flow inversions. 
In this idealized setup, we do not model seismic realization noise, training the focus entirely on the problem of
whether a chosen supergranulation flow model can be seismically recovered. We define the misfit functional as a sum of $L_2$ norm deviations in
travel times between prediction and observation, as measured using short-distance $f$ and $p_1$ filtered and large-distance unfiltered $p$ modes.
FWI allows for the introduction of measurements of choice and iteratively improving the background model,
while monitoring the evolution of the misfit in all desired categories.
Although the misfit is seen to uniformly reduce in all categories, convergence to the true model is very slow, possibly
because it is trapped in a local minimum. The primary source of error is inaccurate depth localization, which,
owing to density stratification, leads to wrong ratios of horizontal and vertical flow velocities (`cross talk').
In the present formulation, the lack of sufficient temporal frequency
and spatial resolution makes it difficult to accurately localise flow profiles at depth. 
We therefore suggest that the most efficient way to discover the global minimum
is to perform a probabilistic forward search, { involving calculating the misfit associated with a broad range of models (generated, for instance, by a Monte-Carlo algorithm) and 
locating the deepest minimum.} Such techniques possess the added advantage
of being able to quantify model uncertainty as well as realization noise (data uncertainty).

%
\end{abstract}

\section{Introduction}
Models of material flows in the interior of the Sun can assist significantly in our understanding of its dynamics.
Consequently, substantial effort has been directed towards seismically imaging flows underneath sunspots \citep[e.g.][]{Duvall1996,zhao, gizonphd,gizon_etal_2009,komm12}, supergranulation \citep{duvall00,beck01,gizon03,zhao03,birch_hanasoge_07,woodard07,gizon08,duvall12,svanda12,duvall14}, 
meridional circulation \citep[e.g.][]{giles97,gizon04,braun08,komm13,zhao13} and convection \citep[e.g.][]{swisdak99,duvall03,hanasoge12_conv,woodard14}. 
Seismic inferences are solutions to inverse problems of the form $Ax = b$, where $b$ is a set of measurements, $A$ is a
matrix comprising of transfer functions between medium properties $x$ and the measurements. Seismology is primarily a methodology
to measure wavespeeds of the medium through which waves propagate. Three types of wavespeeds govern
helioseismic wave propagation \citep{hanasoge12_mag}: an isotropic sound speed (locally independent of direction), a symmetry-breaking, anisotropic flow velocity (locally dependent on the angle of propagation),
and an anistropic (symmetry-conserving) magnetic Alfv\'{e}n velocity. Because flows can break the symmetry of wave propagation,
i.e. waves propagating propagating along and opposite the flow direction are phase shifted in opposite senses, the typical measurement
is of direction-dependent phase-degeneracy lifting, such as $\tau_+ - \tau_-$, where $\tau$ is the travel time and $\pm$ denote pro- and retrograde directions with respect to the flow. 

Measurements of the seismic wavefield of the Sun are made using a number of techniques such as time-distance
\citep{duvall}, ring-diagram analysis \citep{hill88}, holography \citep{lindsey97} and global-mode seismology \citep[e.g.][]{jcd02}. 
The first three techniques are all local in that they are used to infer non-axisymmetric properties of the solar interior \citep[for detailed reviews, see e.g.][]{gizon05,gizon2010}.
The analysis and extraction of seismic data from raw observations, taken for instance by the Helioseismic and Magnetic Imager \citep{hmi}, is 
well understood and the techniques are established. However, the interpretation of these measurements to create models of the solar interior 
has greatly lagged observation. In the context of flow inversions, a number of authors have constructed algorithms \citep[e.g.][]{Kosovichev2000, birch07,jason07,svanda11,hanasoge11,jason12} and performed synthetic tests \citep[e.g.][]{gilesphd,zhao03,hanasoge10,dombroski13,jason14} to verify and validate these techniques.
These validation tests involved calculating the seismic response to an input (user-prescribed) flow system and subsequently inverting the 
responses to test if the original flow system was indeed retrieved.
Unfortunately, the overwhelming fraction these efforts drew the conclusion that seismology was unable to accurately infer the input flow system. 
However, a number of tests were inconsistent since the response to the input flow system was only calculated approximately, using either a
ray approximation or the same sensitivity kernels (transfer functions) that are used in the inversion. Further, because all prior local inversions
have only consisted of one step, non-linearities between model parameters and measurements are not accounted for. Finally, few inversions in the past
have considered satisfying mass conservation of the inverted flows .

A highly successful result in helioseismology is the inference of global rotation shear \citep{schou98}, which has withstood repeated testing and remained
consistent. More complicated flow systems involving lateral and radial flows such as meridional circulation, convection etc. present substantive challenges
in the guise of `cross talk' between vertical and horizontal flows on seismic signatures, making it difficult to distinguish the two. Repeated tests
have shown cross talk to be generally unavoidable \citep{zhao03,dombroski13}. We suggest here that this is due to the poor radial localization
prevalent in non-axisymmetric inversions.

{ The seismic inversion is a projection of a feature (such as 
a supergranule or a sunspot) onto the basis of eigenfunctions of oscillation modes. For global modes (at low-$\ell$), there are a large number
of radial orders $n \lesssim 20$, whereas at relatively high-$\ell$, the regime of interest here, there are far fewer radial orders to choose from (owing to 
the acoustic cutoff frequency which sets the maximum allowed temporal frequency of trapped modes to 5.5 mHz).  Thus the radial resolution is
much finer in global seismology in comparison, inherently placing global inversions on a firmer footing.}

Full Waveform Inversion (FWI), a widely used technique in exploration seismology, is a means of self-consistently solving for model parameters using
characteristics of the entire observed waveform. FWI fundamentally differs from prior classical methods of helioseismic inversion in that it can be iterative, the starting model need not be translationally invariant,
measurements can be introduced at will. The name derives from the goal of fitting the entire waveform by the end of the inversion so that all available seismic information is utilized. It does not necessarily mean that the entire raw waveform is used during the inversion.
Specifically, it is found that parametrising the waveform in terms of classical or instantaneous travel times or amplitudes is an effective strategy towards
fitting the entire waveform
\citep[as opposed to using the raw waveform itself, e.g.,][]{bozdag11, zhu_attenuation, hanasoge2014}. 
In this article we restrict ourselves to classical travel times of well defined parts of the waveform (such as the first or second bounce and filtered times).
Thus the method we are discussing is a subset of a larger collection of techniques termed FWI and hence we refer to it as such.

The first step in the method is to define a cost or misfit functional $\chi$ that comprises the $L_2$ norm of the misfit
between observed and predicted wavefield measurements \citep{hanasoge_tromp}. Since the measured wavefield is a function of the medium of that
waves propagate through, the misfit is effectively a function of model parameters, i.e. $\chi = \chi({\bf m})$, where ${\bf m}$ contains
the medium properties as a function of space. To improve model ${\bf m}$, we consider variations of $\chi$,
\begin{equation}
\chi({\bf m} + \delta{\bf m}) = \chi({\bf m}) + \frac{\partial\chi}{\partial{\bf m}}\cdot\delta {\bf m} + O(|\delta{\bf m}|^2).\label{misfitde}
\end{equation}
Thus if we want $\chi({\bf m} + \delta{\bf m}) < \chi({\bf m})$, then one possible choice is guided by the steepest-descent method: 
$\delta {\bf m} = - \eps\,{\partial\chi}/{\partial{\bf m}}$, where $\eps > 0$ is a small quantity. A faster way to converge to the minimum is
to use a Krylov-subspace technique, and here, we employ the non-linear conjugate-gradient method. 
The local gradient of the misfit functional (its Jacobian) with respect to model parameters, i.e. $\partial\chi/\partial{\bf m}$, is obtained using a computational realization of the
adjoint method \citep[also known as partial-differential-equation-constrained optimization; see,][]{hanasoge11}. { The adjoint method allows for the computational
evaluation of this gradient, also termed broadly as sensitivity kernels.}

Iterative inversions have the benefit that the misfit in a wide variety of categories such as travel times measured in $f$ or $p_1$ or with other phase- and frequency-filtered data, 
can be monitored. A serious drawback of prior flow-inversion testing is that the misfit is never studied post inversion, making the current
approach very attractive. 

Models of the solar interior are functions of space and are high-dimensional quantities. For instance, in the problem considered here, some
120,000 grid points are used to resolve wave propagation and therefore at least as many parameters.  One can therefore consider a {\it distribution} of models, 
described by some probability density function and a given model being one realization drawn from this distribution. 
For each model, there exists a corresponding wavefield which in turn implies one value of the misfit. Thus a high-dimensional quantity is mapped on to one number
and it is the task of inverse theory to converge on the `correct' model that fits the observations. In other words, there is a model that possibly corresponds to a global 
minimum in misfit that we must find. However, there may also be a variety of local minima in this misfit-model space and it is conceivable that a poor initial guess
could lead to the system being trapped in a local minimum. This discussion points to the concept of {\it model uncertainty} implying that in addition to uncertainty in
data (owing to stochastic wave excitation noise), there is a set of models consistent with measurements. 

\begin{figure}[!ht]
\begin{centering}
\epsscale{1.}
\plotone{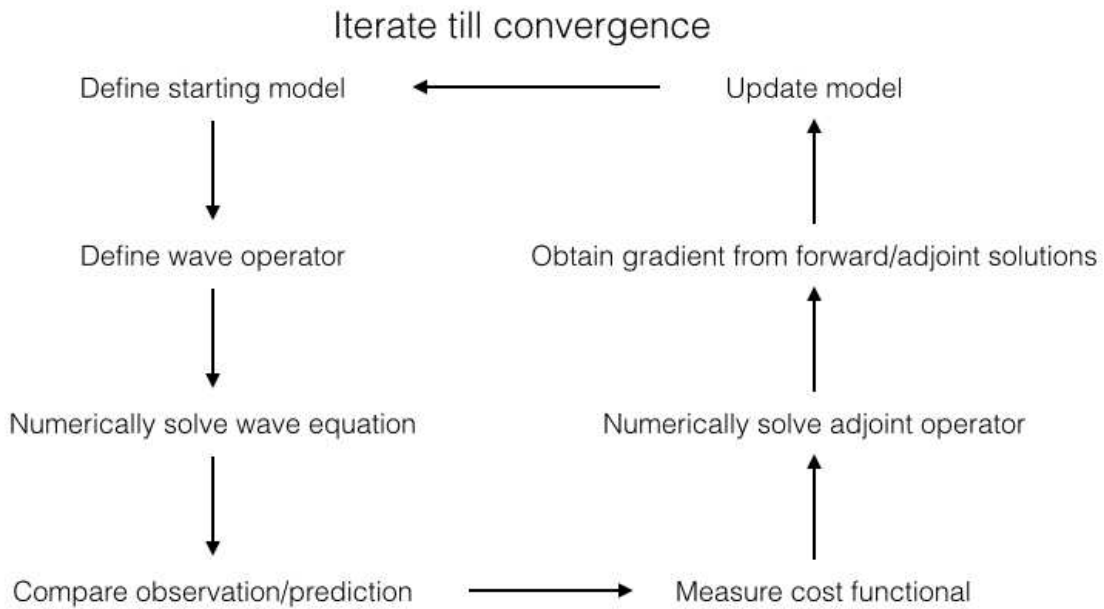}
\caption{Schematic of steps involved in an iterative inversion.
\label{inverseflow}}
\end{centering}
\end{figure}

Inversion strategy (a schematic of which is shown in Figure~\ref{inverseflow}) consists of making a series of choices that limit the likelihood of being trapped in a local minimum. 
This is especially important in exploration seismology where models of the oil reservoir can exhibit strong local heterogeneities. The Sun, a convecting fluid, is
well mixed and consequently, the issue of strong heterogeneities is not a serious issue (with the exception of sunspots) and 
model uncertainty is generally not perceived to be very important.
The acoustic sound speed plays an overwhelmingly important role in wave propagation and therefore, structure inversions have been observed to be 
robust to model uncertainty \citep{hanasoge_tromp}. In other words, for structure-related anomalies, the model-misfit space is such that convergence is likely.
However, we demonstrate here that flow inversions are not as easily tractable. 
Depending on the strategy adopted, i.e., type of measurements
assimilated, preconditioning applied to the kernels etc., a range of models show agreement with measurements and the misfit is seen to smoothly
fall in various categories. 

A commonly used strategy in terrestrial seismology is to separate measurements by frequency and wavelength. Very early on in the inversion, only low-frequency,
large-wavelength modes are used, allowing only coarse changes to model. Once the misfit is sufficiently reduced, then the frequency is increased and relatively
small-wavelength modes are introduced, refining the prior model. This process must be controlled carefully since allowing in small-wavelength modes at the
very start can lead the result down an incorrect path of (misfit) descent. We test the utility of this strategy as well.

\section{Formulation}
{ We start by defining the wave equation that will be studied here. 
Denoting 2-D space by $\bx = (x,z)$, where $x$ and $z$ are the horizontal and vertical coordinates respectively, the equation governing helioseismic wave propagation is
\begin{equation}
\rho\partial^2_t\bxi = 2\rho{\bf v}\cdot\bnabla\partial_t\bxi + \bnabla(\rho c^2\bnabla\cdot\bxi + \rho g\xi_z) +  {\bf g}\,\bnabla\cdot(\rho\bxi) + {\bf S},\label{goveq}
\end{equation}
where $\rho = \rho(\bx)$ is density, ${\bf v} = {\bf v}(\bx)$ is background flow velocity, $c = c(\bx)$ is sound speed, 
${\bf g} = - g(z)\, {\hat{\bf z}}$ is gravity, $\bxi = \bxi(\bx,t)$ is vector wave displacement, whose vertical component is denoted 
by $\xi_z$, ${\bf S} = {\bf S}(\bx,t)$ is the source and $t$ is time. The
spatial gradient is denoted by $\bnabla$ and $\partial_t$ is the partial derivative with respect to time.}

{ In equation~(\ref{goveq}), hydrostatic balance has been already accounted for, which is why background pressure does not
make an explicit appearance. Flows are considered to be small perturbations around this hydrostatic state and are assumed to not
contribute to the force balance. The only parameters that can be varied in equation~(\ref{goveq}) are density, flow velocity and sound speed.}
Variations in the misfit are written therefore in terms of model parameters as
\begin{equation}
\delta\chi = -\int_\odot d\bx\, K_c\,\delta\ln c + K_\rho\,\delta\ln \rho + {\bf K}_\bfv\,\cdot\delta\bfv,\label{misfit}
\end{equation}
where $c$ is sound speed, $\rho$ is density, $\bfv$ is the vector flow velocity and the terms $K_c, K_\rho, {\bf K}_\bfv$
are kernels for these quantities respectively. Because $c, \rho$ are positive-definite quantities, we can study
normalized variations such as $\delta c/c$ and $\delta\rho/\rho$ (and hence the logarithms) and the kernels are
directly comparable. However, because equation~(\ref{misfit}) contains dimensional flow variations $\delta{\bf v}$,
it is  therefore not directly comparable to the other terms. In a constrained-optimization problem where
a variety of terms are competing to explain the misfit, it is important to pose the problem in such a way that all the terms
are dimensionally compatible. In order to do so, let us consider the physics of flow advection and how it phase
shifts waves.

Advection by a flow $\bfv$ induces frequency shifts to a wave with wavevector $\bk$
thus
\begin{eqnarray}
\delta\omega = \bfv\cdot\bk,
\end{eqnarray}
where $\omega$ is the frequency, $\delta$ represents a shift in the respected quantity and $\bk$ is the wave vector.
Defining $\tau$ as the wave travel time, the following approximate relation holds
\begin{eqnarray}
\frac{\delta\tau}{\tau} = -\frac{\delta\omega}{\omega} = -\frac{\bfv\cdot\bk}{c|\bk|} = -\frac{\bfv\cdot\hat{\bk}}{c},
\end{eqnarray}
where $c$ is the sound speed and $\hat{\bk}$ is the normalized wave vector.
Thus the Doppler-shift term in the wave operator is
\begin{eqnarray}
\delta\bl = -2i\omega\rho\delta\bfv\cdot\bnabla,\\
\bnabla\cdot(\rho\bfv) = 0,\label{masscon}
\end{eqnarray}
where constraint~(\ref{masscon}), which enforces mass conservation, must be satisfied.
The gradient of the misfit functional in equation~(\ref{misfitde}) as computed based on the algorithm described in \citet{hanasoge11} is composed of the temporal convolution
between Green's function and its adjoint.
Green's function is the response of the wave operator to a delta source. We denote it by ${\rm G} = G_{ij}(\bx, \bx', \omega)$, where $G_{ij}$ is Green's tensor, $i$ is the direction along which the 
wavefield velocity is measured, $j$ is the direction along which the source is injected, $\bx$ is the receiver and $\bx'$ is the source. Seismic
reciprocity is a statement about the quantity ${\rm G}^\dagger = G_{ji}(\bx',\bx,\omega)$ in relation to the original Green's function. For a system with no flows,
the following statement is true ${\rm G}^\dagger = {\rm G}_{ij}(\bx,\bx',\omega)$ \citep[depending on the boundary conditions;][]{hanasoge11}. However, when there are flows, the statement
is ${\rm G}^\dagger|_{{\bf v}\rightarrow -{\bf v}} = {\rm G}_{ij}(\bx,\bx',\omega)$, which means that the reciprocal Green's function for a system where the flows are reversed in sign is identical
to the original Green's function (with the correct sign of flows). It turns out that maintaining this relationship is critical to self-consistent interpretations of travel times in the Sun.
This relationship is however only valid when mass conservation is maintained. We therefore seek a formulation where mass is explicitly conserved.  

Since we are considering flow inversions in the $x-z$ plane, we may introduce the scalar stream-function $\psi$, 
\begin{eqnarray}
\bfv = \frac{1}{\rho}\bnabla\curl[\rho c\,(\psi - \psi_0)\,\ey],\label{stream}
\end{eqnarray}
where $\psi_0$ is some constant fiducial value whose role is to ensure that $\psi(\bx)$ is a positive definite quantity. When $\psi = \psi_0$, the flow
is identically zero. 
Recalling that the misfit arising from flow perturbations is
\begin{eqnarray}
\delta\chi_{\rm flow} = -\int_\odot d\bx\,{\bf K}_\bfv\cdot\delta\bfv = -\int_\odot d\bx\,{\bf K}_{\bfv}\cdot\frac{1}{\rho}\bnabla\curl\{\delta[\rho c(\psi - \psi_0)]\,\ey\} - {\bf K}_\bfv\cdot\bfv\, \delta\ln\rho.
\end{eqnarray}
Defining
\begin{equation}
K_\psi = \rho c\psi\,\ey\cdot\bnabla\curl\frac{{\bf K}_\bfv}{\rho},
\end{equation}
using the vector identity ${\bf a}\cdot\bnabla\curl{\bf b} = \bnabla\cdot({\bf a}\curl{\bf b}) + {\bf b}\cdot\bnabla\curl{\bf a}$, and noting that we employ
zero-Dirichlet boundary conditions \citep[also see,][]{hanasoge11},
\begin{eqnarray}
\delta\chi &=&  -\int_\odot d\bx\,\delta[\rho c(\psi-\psi_0)]\,\ey\cdot\bnabla\curl \frac{{\bf K}_{\bfv}}{\rho}  - {\bf K}_\bfv\cdot\bfv\, \delta\ln\rho\nonumber\\
&=& -\int_\odot d\bx\,\left[\left(1-\frac{\psi_0}{\psi}\right)\delta\ln\rho + \left(1-\frac{\psi_0}{\psi}\right)\delta\ln c + \delta\ln\psi\right] K_\psi  -{\bf K}_\bfv\cdot\bfv\, \delta\ln\rho.
\end{eqnarray}
The first two terms encode the cross talk between density and flow and between sound-speed and flow.
Redefining kernels for sound speed and density thus
\begin{equation}
K_c \rightarrow K_c + \left(1-\frac{\psi_0}{\psi}\right) K_\psi,\,\,\,\,\,\,\,\,\,\, K_\rho \rightarrow K_\rho + \left(1-\frac{\psi_0}{\psi}\right) K_\psi  - {\bf K}_\bfv\cdot\bfv,
\end{equation}
we arrive at a formulation for the flow inversion that simultaneously satisfies mass conversation and is written in terms of non-dimensional variations
\begin{equation}
\delta\chi = -\int_\odot d\bx\, K_c\,\delta\ln c + K_\rho\,\delta\ln \rho + {K}_\psi\,\delta\ln\psi.\label{remisfit}
\end{equation}

\section{Problem Setup}
We define the `true' flow model of supergranulation based on the formula described by \citet{duvall12} and \citet{duvall14}. We consider no sound-speed or other perturbations. 
The wavefield simulated using this model as measured at the surface of the computational box is termed `data', which we use to perform the inversion.
The wavefield associated with the sequence of flow models in the iterative inversion are termed `synthetics'. 
The goal is to fit synthetics to data by appropriately tuning the flow model. Because we consider neither
density nor sound-speed anomalies in the true model, we invert only for flow perturbations. The inverse problem we are solving is
\begin{equation}
\delta\chi = -\int_\odot d\bx\,  {K}_\psi\,\delta\ln\psi.\label{truemisfit}
\end{equation}
The flow model is 2-D and with no loss of generality, we consider a 2-D inverse problem, along the lines of \citet{hanasoge_tromp}. These reduced
problems place substantially lighter computational demands and provide insight into inversion strategy. 
 
In order to numerically solve equation~(\ref{goveq}), we employ the Cartesian-geometry-based pseudo-spectral solver {\rm{SPARC}} developed by \citet{dealias} and \citet{Hanasoge_couvidat_2008}.
All spatial derivatives are computed using a sixth-order accurate, compact-finite-difference scheme \citep{lele92}.
An optimized second-order five-stage Runge-Kutta technique \citep{hu} is used to evolve the equation in time. We place perfectly matched layers \citep{hanasoge_2010} on the side and vertical boundaries in order to absorb outgoing waves.

The cost functional used in these calculations is the $L_2$ norm difference between predicted and observed wave travel times $\tau$
at a number of spatial locations $i$ on the surface,
\begin{equation}
\chi = \frac{1}{2}\sum_i (\tau_i - \tau^{\rm o}_i)^2,
\end{equation}
where $\tau_i$ is predicted and $\tau^{\rm o}_i$ is observed. The goal is to minimise $\chi$ knowing that $\tau_i = \tau_i(\bxi)$ { and because
the misfit is dependent on background model parameters ${\bf m}$, $\tau_i(\bxi) = \tau_i({\bf m})$. Thus the idea is to carefully follow the nested
dependencies of wavefield measurements to eventually make the connection to model parameters (i.e. flows in this case).}

\subsection{Full Waveform Inversion}\label{proced}
\begin{figure}[!ht]
\begin{centering}
\epsscale{1}\vspace{-1cm}
\plotone{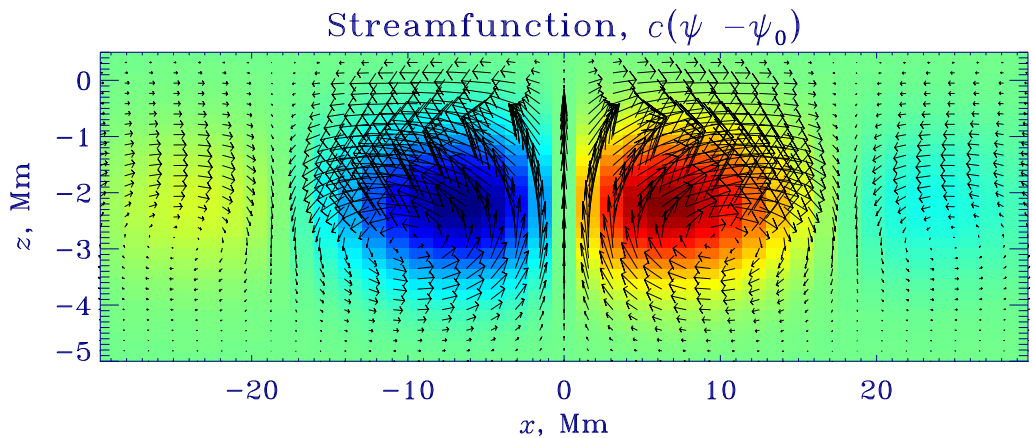}\vspace{-1.2cm}
\caption{True flow model. The contours show the stream function and arrows indicate the true velocity profile (see Eq.~[\ref{stream}]). 
The longest arrow represents a flow speed of 600 m/s. The arrows along the centre line are difficult to discern but the maximum vertical flow occurs at $x=0$, of order 250 m/s.
The prescription for this flow is taken from the mass-conserving model discussed by \citet{duvall12}.
\label{trueflow}}
\end{centering}
\end{figure}

\begin{figure}[!ht]
\begin{centering}
\epsscale{1.}
\plotone{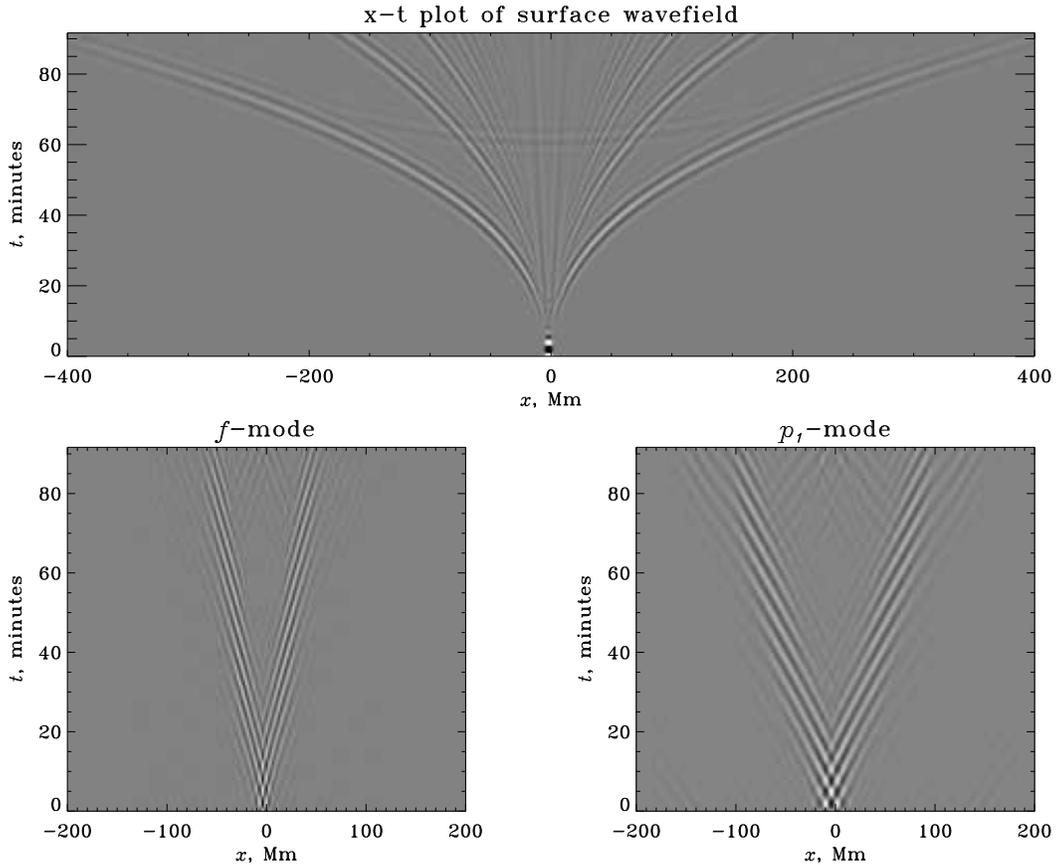}
\caption{Time-distance plot of the wavefield observed at the surface. The source is placed at $x=-1$ and 
waves that are generated are observed at various distances and times at the surface. The lower plots shows
$f$ and $p_1$-mode-filtered surface wavefields. { The parabolic feature in the top panel is due to the partial reflection
of waves from the bottom computational boundary. This occurs despite the utilization of ostensibly high-fidelity-absorptive perfectly matched layers.
Because we use temporal Fourier transforms, the signal wraps around in time as seen in the bottom two panels. This causes some `noise' in the
measurements at late times but is unlikely to play a serious role because the overlap is minor.}
\label{xt}}
\end{centering}
\end{figure}

FWI comprises techniques widely used in geophysics \citep[and recently in helioseismology; ][]{hanasoge_tromp} 
to make self-consistent inferences of complex heterogeneities in Earth's interior. The following summarises the steps for the test problem studied here:
\begin{itemize}
\item Construct a true model of the flow, Figure~\ref{trueflow}, and compute the wavefield at the surface (which we shall term `observations' here),
\item Choose a set of optimally placed sources and a broad set of receivers, since the computational expense scales with the number of sources \citep{hanasoge_tromp},
\item Determine the surface wavefield for a given model of the solar interior using equation~(\ref{goveq}) and compute the predicted wavefield (the forward calculation), 
as shown in Figure~\ref{xt},
\item Choose which measurements to use in the inversion: low-frequency, large-wavelength modes at the start, gradually
introducing higher-frequency data,
\item Compute the misfit between predicted and observed data,
\item { Sum over the gradients (kernels) between every source-receiver pair weighted by the associated travel-time misfit},
\item Compute the gradient of the misfit with respect to the model parameters using the algorithm described in \citet{hanasoge11},
\item Perform a line search to determine the update that results in the greatest misfit reduction,
\item Update the model and repeat.
\end{itemize}


\subsection{Supergranulation}
We use a 2D Cartesian computational grid with $512\times300$ points, spanning $800\times138 \,{\rm Mm}^2$ in the horizontal and
vertical directions respectively. The vertical grid is uniform in acoustic travel time, extending from $r=0.8R_\odot$ to $r=1.001R_\odot$.
We place the supergranule model in Figure~\ref{trueflow} at the horizontal centre of the domain. We choose seven sources (or master pixels)
at locations across the supergranule and wavefield measurements at hundreds of receivers on either side of the source are used in the
inversion. Sources are at a fixed radial location of 100 km below the photosphere and receivers are placed 200 km above, mimicking the
excitation and measurement processes in the Sun. We do not fully model cross-correlation measurements in this work because of the additional
added complexity. Instead, we limit ourselves to deterministic sources, much as in \citet{hanasoge_tromp}, since our aim is to establish the
viability of FWI for flow inversions. { In order to start with no bias and conditions similar to the inversion in \citet{hanasoge_tromp},}
we start with model $\psi = \psi_0$, i.e. no flows. Subsequently, we follow the procedure outlined in Figure~\ref{inverseflow} and
section~\ref{proced}. For measurements, we use a continuous span of receivers, starting with $f$-filtered travel times for source-receiver distances 
ranging from 8 to 25 Mm, $p_1$-filtered travel times for distances from 10 to 35 Mm and first-bounce unfiltered $p$ travel times for distances from
35 to 350 Mm (see Figure~\ref{xt}). { We recognize that it is important to include spherical-geometric effects for such substantial wave travel distances
when dealing with observations. However, in this case, the `data' come from the same numerical code, so the approach is consistent. To reiterate, we
use large-distance measurements to test their efficacy in improving the fidelity of inversions.}
Frequency filters are not applied although some minor experimentation showed benefits to be limited. 
We precondition the gradient with the approximate Hessian described in \citet{tromp13} and \citet{zhu_attenuation}.

\begin{figure}[!ht]
\begin{centering}
\epsscale{1.}
\plotone{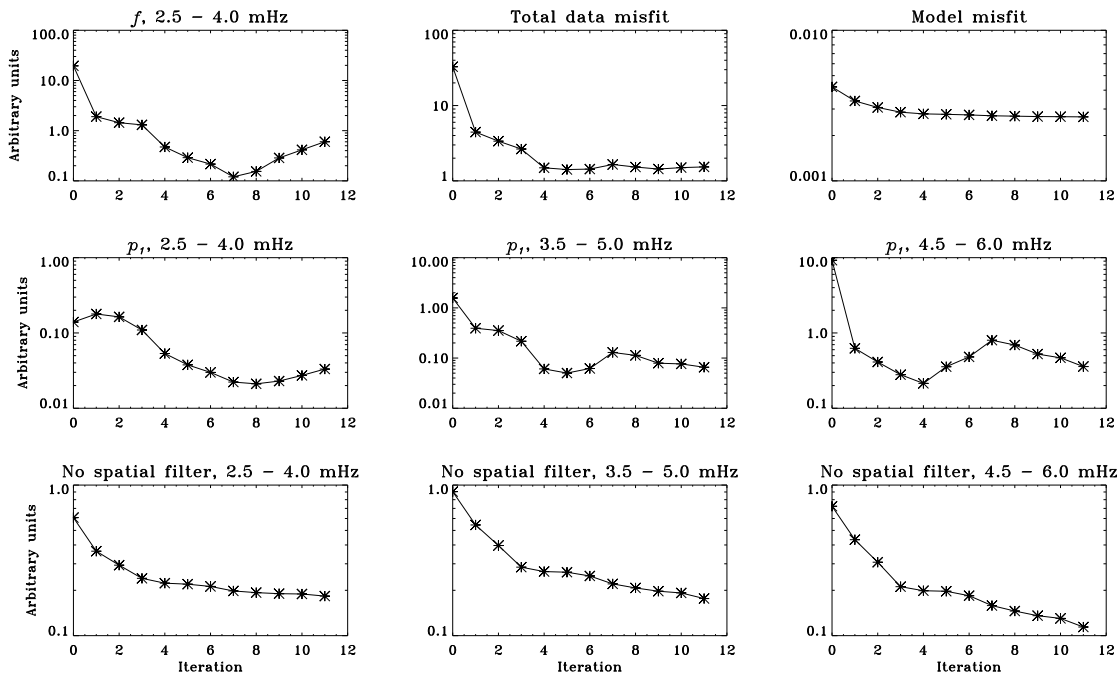}
\caption{Misfit as a function of category and iteration. The $f$-mode misfit falls by a factor of almost hundred before rising again,
trading off with the $p_1$ and $p$-mode misfit. { We note that we only use large-distance measurements when estimating
the misfit of the lowest panels; for such large distances, the signal is entirely comprised by $p$-modes}. 
The total data misfit changes very slowly beyond the first few iterations, a manifestation of the tradeoff between different modes.
The model misfit, which is defined as the $L_2$ norm of the difference between
the true and inverted models is seen to decrease very slowly.
\label{misfitfig}}
\end{centering}
\end{figure}

\begin{figure}[!ht]
\begin{centering}
\epsscale{1.2}
\plotone{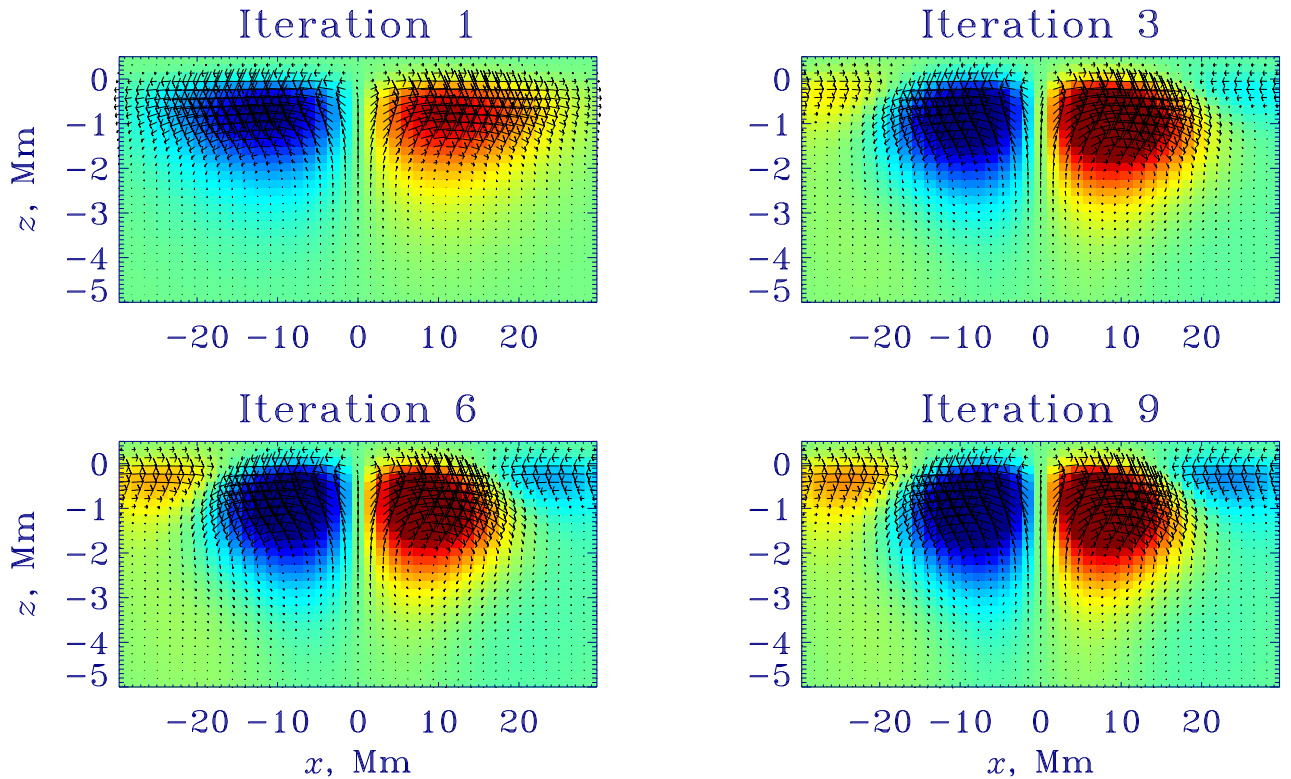}
\caption{Flow model as it evolves with iteration. The flow model gradually converges to
the true model but at a very slow rate. The iterated models peak at the surface whereas the true model (Figure~\ref{trueflow})
peaks at a deeper layer. In the current inversion, errors in vertical and horizontal flow speeds occur due to poor depth
localization.
\label{iterflow}}
\end{centering}
\end{figure}

\begin{figure}[!ht]
\begin{centering}
\epsscale{1.2}
\plotone{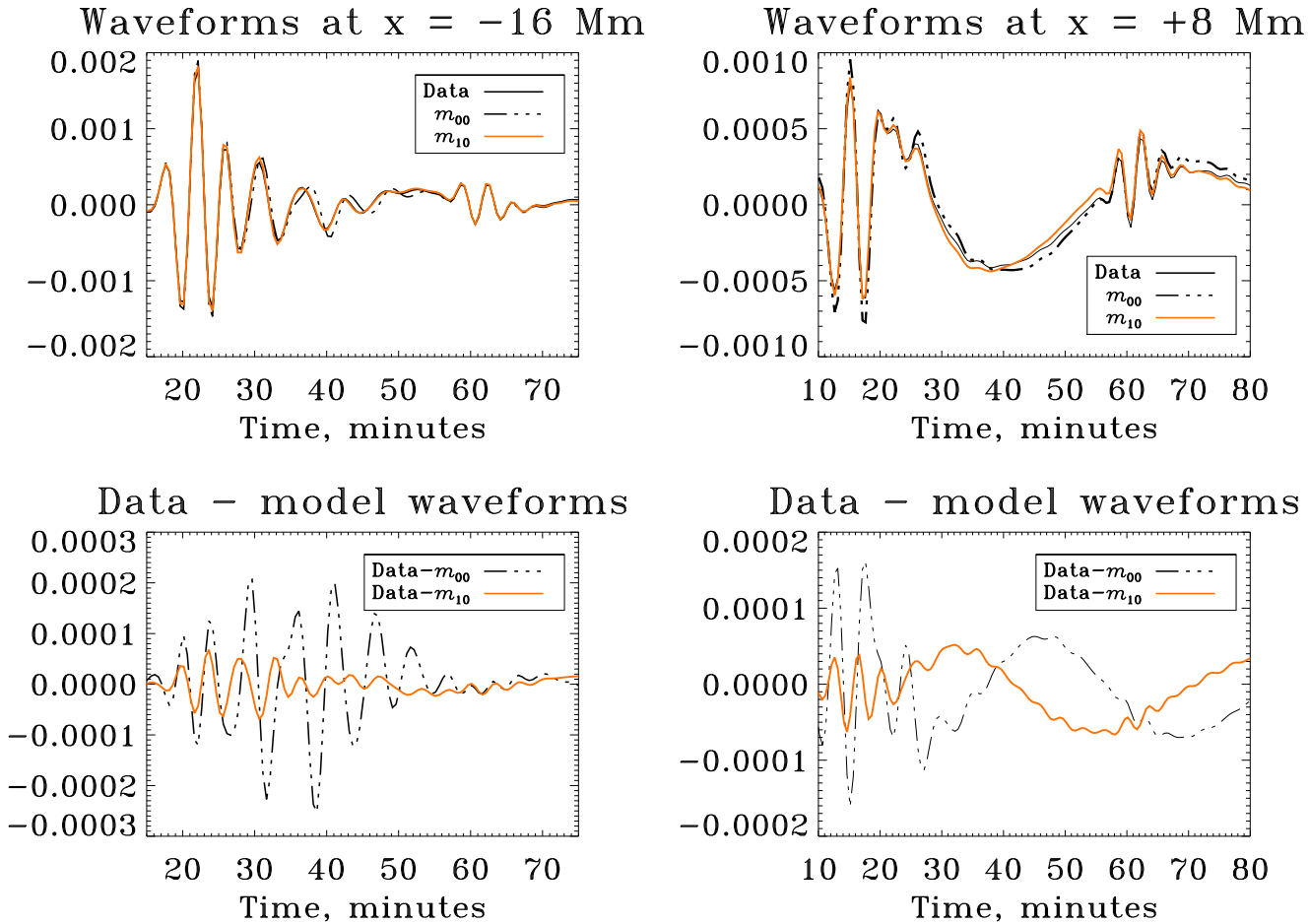}
\caption{Evolution of the waveform with iteration where the source is at $x=-1$ Mm. Shown are `data' and waveforms for models at iterations 0 and 10 recorded at different $x$ locations on the surface. Essentially these are cuts at constant $x$ of the time-distance plot (Figure~\ref{xt}). Because the data and model 10 waveforms are difficult to distinguish in the upper panels, we show the difference between data and models at iterations 0 and 10 in the bottom two panels for the same $x$ locations. The waveform misfit is seen to reduce but not as significantly as one might hope.
\label{xtdiff}}
\end{centering}
\end{figure}

As advertised, and displayed in Figure~\ref{misfitfig}, the misfit decreases uniformly in every measured category, at least for the first several
iterations. The evolution of the model of the supergranule is shown in Figure~\ref{iterflow} and the raw waveform misfit in Figure~\ref{xtdiff}.
The model is strongly surface peaked, and improvements at depth occur very slowly, and we find that kernels for the inversion have
concentrated power in the near-surface layers. A careful examination of the $f$, $p_1$ and $p$ kernels (not shown here) reveals that
it is difficult to remove the effect of the surface \citep[reminiscent of the `shower-glass' effect;][although that was applied to magnetic fields]{schunker05}. 
Beyond a certain number of iterations, Figure~\ref{misfitfig} illustrates a tradeoff between $f$, $p_1$ and $p$ modes, signalling
incorrect depth localization of the flow model. 

It is worth considering why global helioseismic inversions have been successful in inferring rotation, since, ostensibly, global modes
have similar systematical biases. A substantial advantage of global modes is their high resolution in radial order and spherical harmonic
wavenumber, allowing for the direct manipulation of global-mode kernels in order to diminish the surface tail. In our analysis, we treat
travel times obtained from the seismic waveform using its entire available bandwidth, i.e. we do not apply frequency filtering
(separating measurements in frequency appears to have only minor gains). This results in a diminished frequency resolution in
comparison to global modes, contributing to the poor localization in depth.

Nevertheless, these poor convergence properties are in strong contrast with the structure inversions
performed by \citet{hanasoge_tromp}, also using FWI. In this prior work, \citet{hanasoge_tromp}, with a smaller set of measurements, 
were able to recover the details of a sound-speed perturbation that they had inserted. One may speculate that flow inversions possess
a larger null space but whether this holds water remains to be determined. It is likely that the seismic measurements contain
sufficient information to discern between models peaked at different depths, but that the inversion is poorly conditioned, resulting in
low convergence rates. We therefore suggest a probabilistic forward search as a means of locating the global minimum.

\section{Discussion and Conclusions}
We have formally studied the problem of inferring subsurface flows of material in the Sun. 
Rotation stands unique as a well tested and verified flow system. However rotation
is an entirely lateral flow and is axisymmetric. Flows with overturning motions such as meridional
circulation, supergranulation and convection show radial and lateral motions, the latter 
being non-axisymmetric. Further, meridional circulation is antisymmetric across the equator, making
it a difficult target to image using global modes \citep[although unconventional means have been explored by][]{woodard13}.
Consequently, local targeting methods such as time-distance, ring analysis and holography are necessary.
It has however been pointed out that in overturning mass flows, the vertical and horizontal components
are not easily distinguished \citep{zhao03}, and therefore it is unclear if even local methods are able to overcome this issue. 

In the current work, we explore the question of whether it is possible to infer a (known) flow, in the ideal limit of zero realization noise.
We simulate wave propagation through the true model, which in this case is the supergranulation flow model from \citet{duvall12}, shown in Figure~\ref{trueflow},
terming the surface wavefield measurements as `data'. Starting from
the quiet Sun (i.e. no background flow), we apply full waveform inversion, a technique where the flow model is iteratively
improved. The inversion is constrained by the governing wave equation with mass conservation being strictly enforced.
The misfit functional for this problem is defined as the $L_2$ norm of the difference between observed and predicted travel times.
We monitor the variation of misfit with iteration in a variety of categories, applying ridge and frequency filters to the surface wavefield.
The measurements that are actually used in the inversion are not frequency filtered, so it is interesting to note that the misfit in
these categories reduces independently.
It is seen that the misfit uniformly reduces in all categories (e.g. see Figure~\ref{misfitfig}) but the flow appears to converge very slowly
to the true model (see Figure~\ref{iterflow}). 

We tried two different strategies, one in which all measurements are introduced at the start of inversion and another where
only large-distance $p$-mode travel times are introduced at the first iteration and subsequently, $f$ and $p_1$ modes are
added. The latter appeared to converge somewhat more rapidly, but neither strategy produced the correct model. The basic
error in the model is the inaccuracy in recovering the vertical flow velocity, which was almost a factor of ten smaller then the
true model. This occurred because the depth of the inverted flow was not correctly obtained. At the end of the inversion,
the overall misfit fell by over a factor of 10 but saturated at this level. Further iterations did not cause the misfit to decrease appreciably, 
resulting rather in a misfit tradeoff between $f$ and $p_1$. Thus appears to suggest that the inversion is trapped in some sort of local 
minimum. { The issue may be traced to the fact that for small-scale features, there are relatively few modes that can be
used in the seismic analysis (at high-$\ell$, the power spectrum shows a sparsity in number of modes). Thus resolving the depth
structure of these features is difficult. A corollary to this insight is that it is not evident that even by starting at a model that is very close to the 
true model, the inversion will push the model towards to the right direction. \citet{duvall12} introduced the idea of using large-distance
measurements; such modes ostensibly increase the number of radial orders available to the inversion. However, even this set of measurements
was found to be insufficient in the end.
In contrast, it must be noted that \citet{hanasoge_tromp} were, with reasonable accuracy, able
to recover a thermal structure anomaly using FWI, whereas the flow inversion here has not been nearly as successful. That reason
for this dichotomy is not yet fully apparent but one possibility is that flow inversions also have a serious null-space issue. It is unlikely that
increasing the number of observations will improve the convergence rate.}

Despite the elaborate nature of this technique, the inversion failed converge to the correct solution. 
In contrast, conventional, local flow inversions involve only one step, and no independent means of verifying 
that the flow model explains the seismic measurements better are applied.
It is therefore important to estimate the model uncertainty by studying the 
misfit associated with a class of models. { A probabilistic search over a range of forward models, \`{a} la \citet{khan09} for instance, which involves simulating waves
through a number of models, measuring the misfit associated with each and locating the deepest minimum, is
a useful technique. Additionally, this removes the limitation of trying to project the flow model on the limited set of eigenfunctions available at high wave numbers
and one can use a larger variety of models \citep[also e.g.,][]{cameron08}.}
Despite the large number of surface observations,  
it is surprising to note that it is not just realization noise (data uncertainty) that determines the accuracy of imaging flows in the solar interior
but also likely model uncertainty. 

The implications for the supergranulation models of \citet{duvall12}, \citet{svanda12} and \citet{duvall14}, which involve large-magnitude
vertical and horizontal flows is not entirely clear. While \citet{duvall12} suggest the use of large-distance measurements, which correspond
to large-wavelength (coarse-scale) modes, it is not apparent that the eventual solution is accurate. 


The problem could also be posed in alternate formulations, such as by connecting the flow
to the stream function in equation~(\ref{stream}) according to
\begin{equation}
\bfv = \frac{1}{\rho}\bnabla\curl[\rho c^2\,(\psi - \psi_0)\,\ey],
\end{equation}
which places greater weight on the inversion in deeper layers ($\propto c^2$). We attempted this approach and while this produces larger
vertical velocities, the solution had moved to a different local minimum (and not the global minimum). 
However this avenue remains to be explored more thoroughly. 

Admittedly, this result is discouraging in that even in this idealized inversion, the flow cannot be exactly recovered. Thus, model uncertainty
should be considered as a critical part of the inversion. 
A forward search over a broad class of flow models may be the most productive 
technique since this would, in addition to potentially discovering the global minimum, allow us to map the model-misfit space. 
We could then place model uncertainties on the flow inversion, which together with data uncertainty or realization noise, would allow for more accurate
uncertainty quantification. 

\acknowledgements
The computation for these results was performed using NASA's Pleiades cluster. SMH thanks the Indian government for funding this research
and J. Schou, H. M. Antia and A. C. Birch for useful conversations.


\bibliographystyle{apj}
\bibliography{ms}

\end{document}